\documentclass[prl,aps,twocolumn,showpacs]{revtex4}
\usepackage{graphicx}
\usepackage{amsmath}
\usepackage{bm}
\usepackage{braket}
\usepackage{amstext}
\usepackage{amsxtra}
\usepackage{float}
\usepackage{upgreek}

\usepackage{times}

\newcommand{\kB}{k_{\rm B}}

\newcommand{\mJT}{\mu_{\rm JT}}
\newcommand{\cJT}{c_{\rm JT}}

\begin{document}

\title{Quantum Joule-Thomson Effect in a Saturated Homogeneous Bose Gas}

\author{Tobias F. Schmidutz$^\star$, Igor Gotlibovych$^\star$, Alexander L. Gaunt, Robert P. Smith, Nir Navon$^\dag$, and Zoran Hadzibabic}
\affiliation{Cavendish Laboratory, University of Cambridge, J. J. Thomson Avenue, Cambridge CB3 0HE, United Kingdom }

\begin{abstract}
We study the thermodynamics of Bose-Einstein condensation in a weakly interacting quasi-homogeneous atomic gas, prepared in an optical-box trap. 
We characterise the critical point for condensation and observe saturation of the thermal component in a partially condensed cloud, in agreement with Einstein's textbook picture of a purely statistical phase transition.
Finally, we observe the quantum Joule-Thomson effect, namely isoenthalpic cooling of an (essentially) ideal gas. In our experiments this cooling occurs spontaneously, due to energy-independent collisions with the background gas in the vacuum chamber.  We extract a Joule-Thomson coefficient $\mJT > 10^9$~K/bar, about ten orders of magnitude larger than observed in classical gases.
\end{abstract}

\date{\today}


%03.75.Hh	Static properties of condensates; thermodynamical, statistical, and structural properties
%03.75.Nt	Other Bose-Einstein condensation phenomena
%05.30.-d	  	Quantum statistical mechanics
%67.85.-d 	Ultracold gases, trapped gases
%03.75.Kk 	Dynamic properties of condensates; collective and hydrodynamic excitations, superfluid flow
%42.50.Tx	Optical angular momentum (quantum optics)
%47.37.+q 	Hydrodynamic aspects of superfluidity; quantum fluids
%67.85.De 	Dynamic properties of condensates; excitations, and superfluid flow
%37.10.Vz 	Mechanical effects of light on atoms, molecules, and ions
%37.10.De	Atom cooling methods
%37.10.Gh	Atom traps and guides
%37.10.Jk	Atoms in optical lattices
%67.85.Hj		Bose-Einstein condensates in optical potentials

\maketitle

Ultracold atomic gases provide textbook-like demonstrations of basic quantum-statistical phenomena, such as Bose-Einstein condensation~\cite{Anderson:1995} and Fermi pressure~\cite{Truscott:2001},
and allow controllable studies of open problems in the physics of strongly correlated interacting systems~\cite{Bloch:2008}.
However, they are traditionally produced in harmonic traps, which makes them spatially inhomogeneous and thus different  both from other correlated systems, that they often aim to emulate, and from textbook models.
This difference can often be addressed using local density approximation (LDA), and is in some cases even beneficial~\cite{Ho:2010a, Nascimbene:2010, Navon:2010, Hung:2011, Yefsah:2011, Smith:2011b, Ku:2012}. 
However, spatial inhomogeneity is also often problematic. Integrating over the varying density can smear or even qualitatively change experimental signatures (see, e.g., ~\cite{Raman:1999, Perali:2011,Nascimbene:2011, Tammuz:2011, Smith:2011}). 
Moreover, LDA breaks down close to phase transitions, where the correlation length diverges~\cite{Donner:2007}.
Recently, a Bose-Einstein condensate (BEC) of atoms in an essentially uniform potential was achieved~\cite{Gaunt:2013}, opening new possibilities for studies of both quantum-statistical and interaction effects in many-body systems.

\begin{figure} [bp]
\includegraphics[width=0.9\columnwidth]{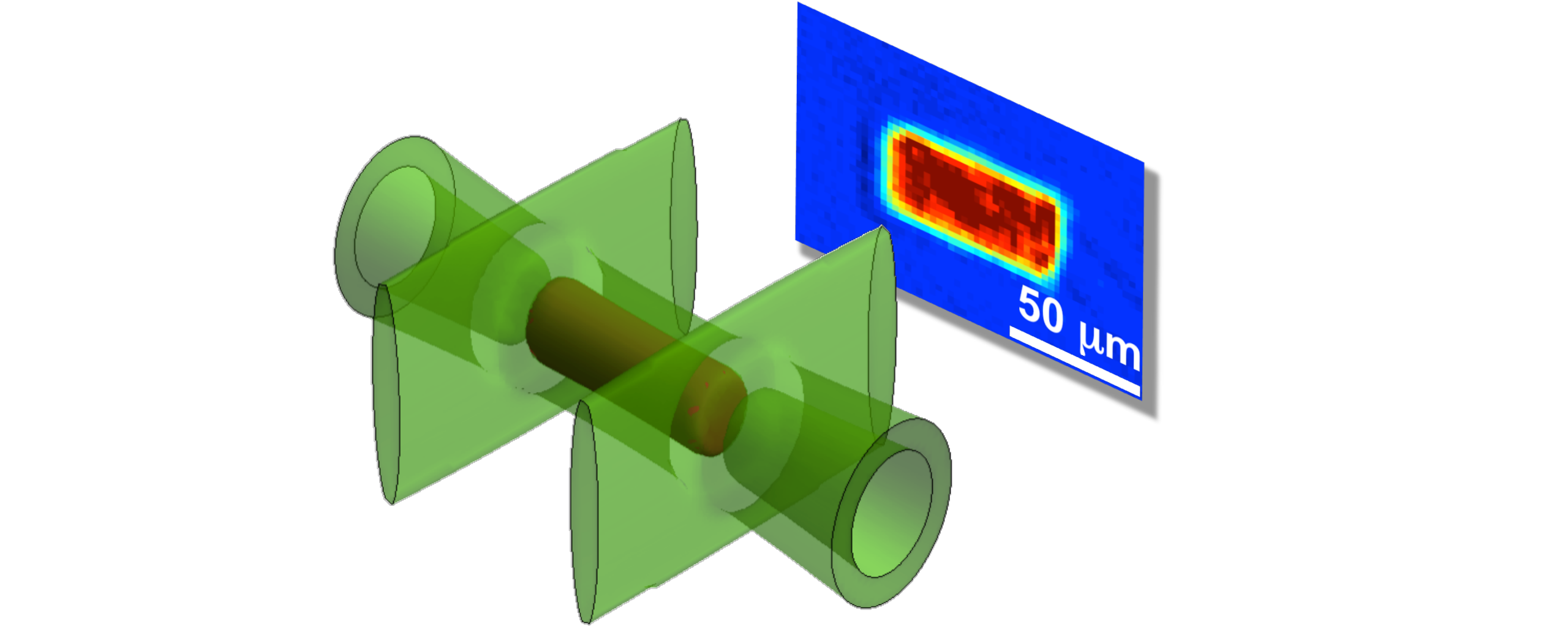}
\caption{(color online) Quasi-homogeneous Bose gas in an optical-box trap. 
We show a schematic of our cylindrically-shaped trap, formed by intersecting one tube-like and two-sheet like green laser beams, and 
an in-situ absorption image of the atomic cloud.}
\label{fig:box}
\end{figure}

In this Letter, we explore the thermodynamics of a weakly interacting quasi-homogeneous Bose gas, prepared in an optical-box trap (see Fig.~\ref{fig:box}). We characterise the critical point for condensation and demonstrate two purely quantum-statistical phenomena that are, despite weak interactions, obscured in harmonically trapped gases.
First, we observe saturation of the thermal component in a partially condensed gas, in agreement with Einstein's textbook picture of condensation as a purely statistical phase transition. 
Second, we observe and quantitatively explain cooling of partially condensed clouds through isoenthalpic rarefaction, driven by collisions with the background gas in the vacuum chamber. This phenomenon is the quantum version of the Joule-Thomson (JT) effect, exploited in
thermal machines such as refrigerators and heat pumps. In the classical JT process, isoenthalpic cooling occurs {\it only} due to interactions, whereas a quantum Bose gas is expected to show it even in their absence. This phenomenon was predicted as early as 1937~\cite{Kothari:1937}, and was at the time proposed as a way to experimentally observe the quantum statistics obeyed by a gas, the effect being opposite for fermions (see also~\cite{Timmermans:2001a}).
However, to our knowledge, JT cooling has never before been experimentally observed in the regime where quantum-statistical effects dominate over the interaction ones.
Outside the realm of ultracold atomic gases, the quantum JT effect is many orders of magnitude smaller~\cite{Kothari:1937}, whereas in ultracold-atom experiments it is usually suppressed because of harmonic trapping and the resulting non-saturation of the thermal component. 

Our experimental setup is described in detail in~\cite{Gotlibovych:2013,Gaunt:2013}. 
We produce BECs of $^{87}$Rb atoms in the quasi-uniform potential of the optical-box trap depicted in Fig.~\ref{fig:box}.
The repulsive walls of the cylindrically-shaped box are created by one tube-like and two sheet-like 532~nm laser beams, and we use a magnetic field gradient to cancel the linear gravitational potential.
The box trap is loaded from a cloud pre-cooled to $T \sim 100$~nK in a harmonic trap. Further cooling is achieved by forced evaporation in the box potential.

In a box trap, the BEC spreads over the same volume as the thermal cloud, making the condensate density and the total interaction energy much lower than in harmonically trapped gases with similar atom numbers, temperatures and s-wave scattering length ($a\approx 5$~nm for $^{87}$Rb).
In all our measurements, with condensed fractions up to $40\%$, the interaction energy is $\lesssim 10\%$ of the thermal energy. 
We therefore expect the thermodynamics of our clouds to be well described by ideal-gas theory.

\begin{figure} [tp]
\includegraphics[width=\columnwidth]{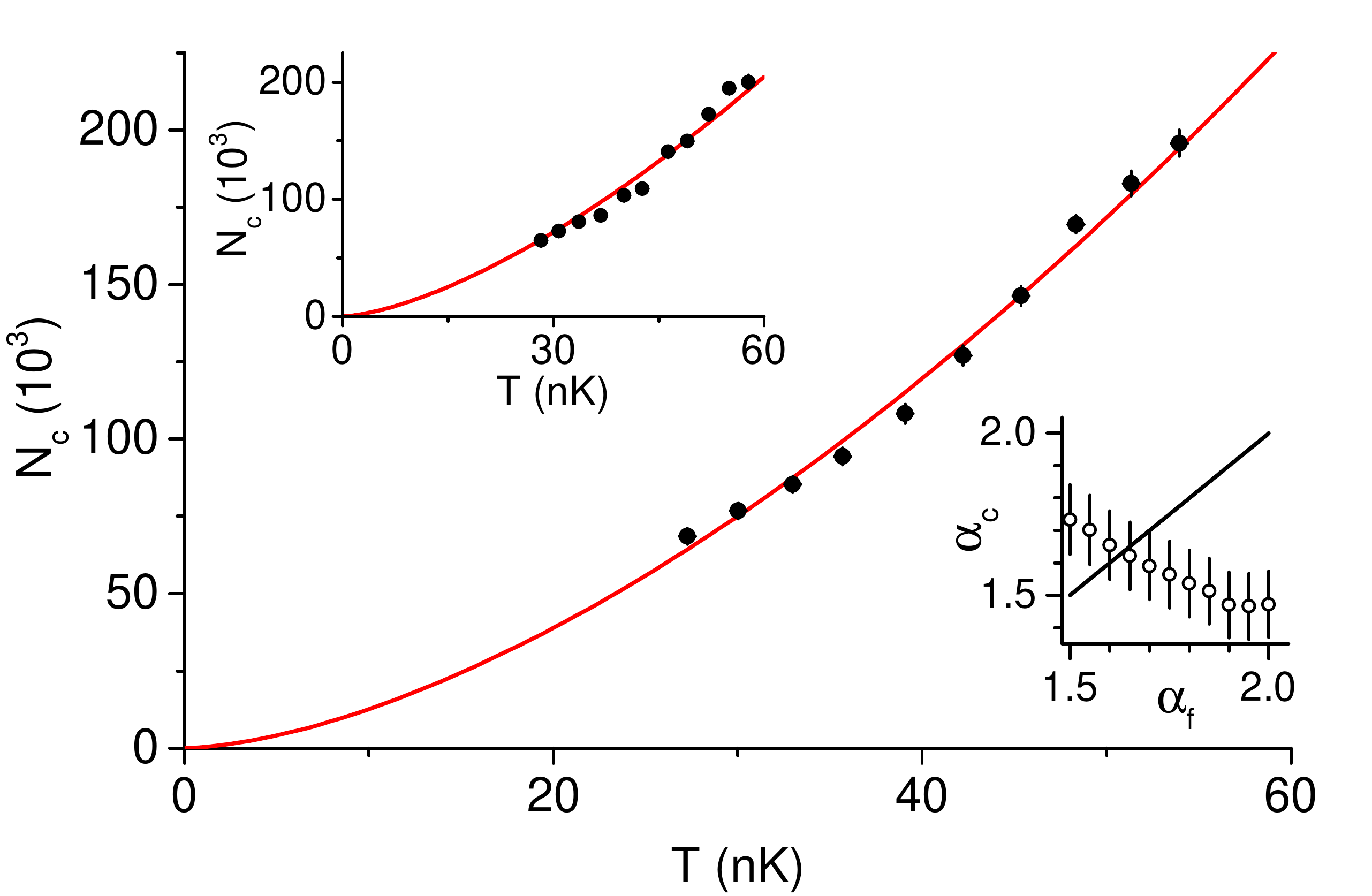}
\caption{(color online) Critical point for condensation. 
A power-law fit, $N_c \propto T^{\alpha_c}$, gives $\alpha_c \approx \alpha_f = 1.65$ (see text).  Bottom inset: within errors, $\alpha_c = \alpha_f$ (solid line) is satisfied for $1.6 \lesssim \alpha_f \lesssim 1.7$. Top inset: comparison with theory of a perfectly homogeneous gas, with fixed $\alpha_c = \alpha_f = 3/2$. }
\label{fig:Nc}
\end{figure}

We first study the critical atom number for condensation, $N_c(T)$, which also allows us to quantitatively characterise our trap (see Fig.~\ref{fig:Nc}). Due to diffraction effects, the optical trap walls cannot be infinitely steep. The resulting small deviation from a perfectly uniform box can to leading order be modelled by an effective high-power-law potential, $\propto r^n$~\cite{Gaunt:2013}. Defining $\alpha=3/2 + 3/n$, the thermal momentum distribution is given by the polylog function $g_{\alpha-3/2}$, the density of states is  $\rho(\varepsilon)\propto \varepsilon^{\alpha-1}$, and we expect~\cite{Pethick:2002}
\begin{equation}
N_c \propto T^{\alpha} .
\label{eq:Nc_trap}
\end{equation}

To measure $T$ and the number of thermal ($N'$) and condensed ($N_0$) atoms, we image clouds after $50-70$~ms of time-of-flight (ToF) expansion from the trap.
We vary $T$ by changing the evaporation sequence, and the total atom number $N$ by changing the initial loading of the trap. After forced evaporation we always raise the trap depth to $U_0 \sim \kB \times 0.4\,\mu$K, where evaporation is negligible, and hold the gas for another 1~s before ToF.
We identify the critical point with the appearance of small BECs, with $N_0 \lesssim 5000$~\cite{Nc}. 

Using Eq.~(\ref{eq:Nc_trap}) to determine the $\alpha$ value for our trap requires some care, because extracting $N_c(T)$ from ToF images relies on a polylog fitting function that itself depends on $\alpha$. To ensure our analysis is self-consistent, we proceed as follows: We first fit all images using various $\alpha$ values, $\alpha_f$, and get a separate $N_c(T)$ curve for each $\alpha_f$. Then, fitting these curves to Eq.~(\ref{eq:Nc_trap}), we get $\alpha_c$ for each $\alpha_f$. Finally, requiring $\alpha_c = \alpha_f$ for self-consistency, we get $1.6 \lesssim \alpha \lesssim 1.7$; see main panel and bottom inset of Fig.~\ref{fig:Nc}~\cite{Old_alpha}.  For the data analysis in Figs.~\ref{fig:saturation} and \ref{fig:JT} below, we fix $\alpha_f=1.65$, corresponding to $n=20$. 

In a perfectly uniform potential $N_c \propto V T^{3/2}$, where $V$ is the box volume. Small deviation of $\alpha$ from $3/2$ corresponds to a slight increase of the effective trapping volume with temperature, $V \propto T^{\alpha - 3/2}$, due to the non-infinite steepness of the trap walls. 
From in-situ images (as in Fig.~\ref{fig:box}), we measure a $(20\pm10)\%$ increase in $V$ between $T=25$ and $50$~nK, which is consistent with our $\alpha$. 

Over the whole temperature range in Fig.~\ref{fig:Nc}, for the critical phase space density we get $\rho_c = N_c \lambda^3/V = 2.0 \pm 0.2$, where $\lambda = \hbar \sqrt{2\pi/ m \kB T}$ is the thermal wavelength, and $m$ is the atomic mass.
Theoretically we expect $\rho_c \approx 2.4$, including finite-size correction~\cite{Grossmann:1995} to the thermodynamic-limit $\rho_c \approx 2.612$~\cite{Pethick:2002}. This small discrepancy is within our systematic uncertainties of $10\%$ in $N_c$ and $20\%$ in $V$~\cite{systematics,disorder}.

While in the above we were careful to characterise the small deviation of our trap from an ideal box, we note that simply assuming a perfectly homogeneous gas of constant volume leads to very small errors. This is shown in the top inset of Fig.~\ref{fig:Nc}, where we fix both $\alpha_f$ and $\alpha_c$ to $3/2$ and the fit of $N_c(T)$ is still very good.

\begin{figure} [tbp]
\includegraphics[width=\columnwidth]{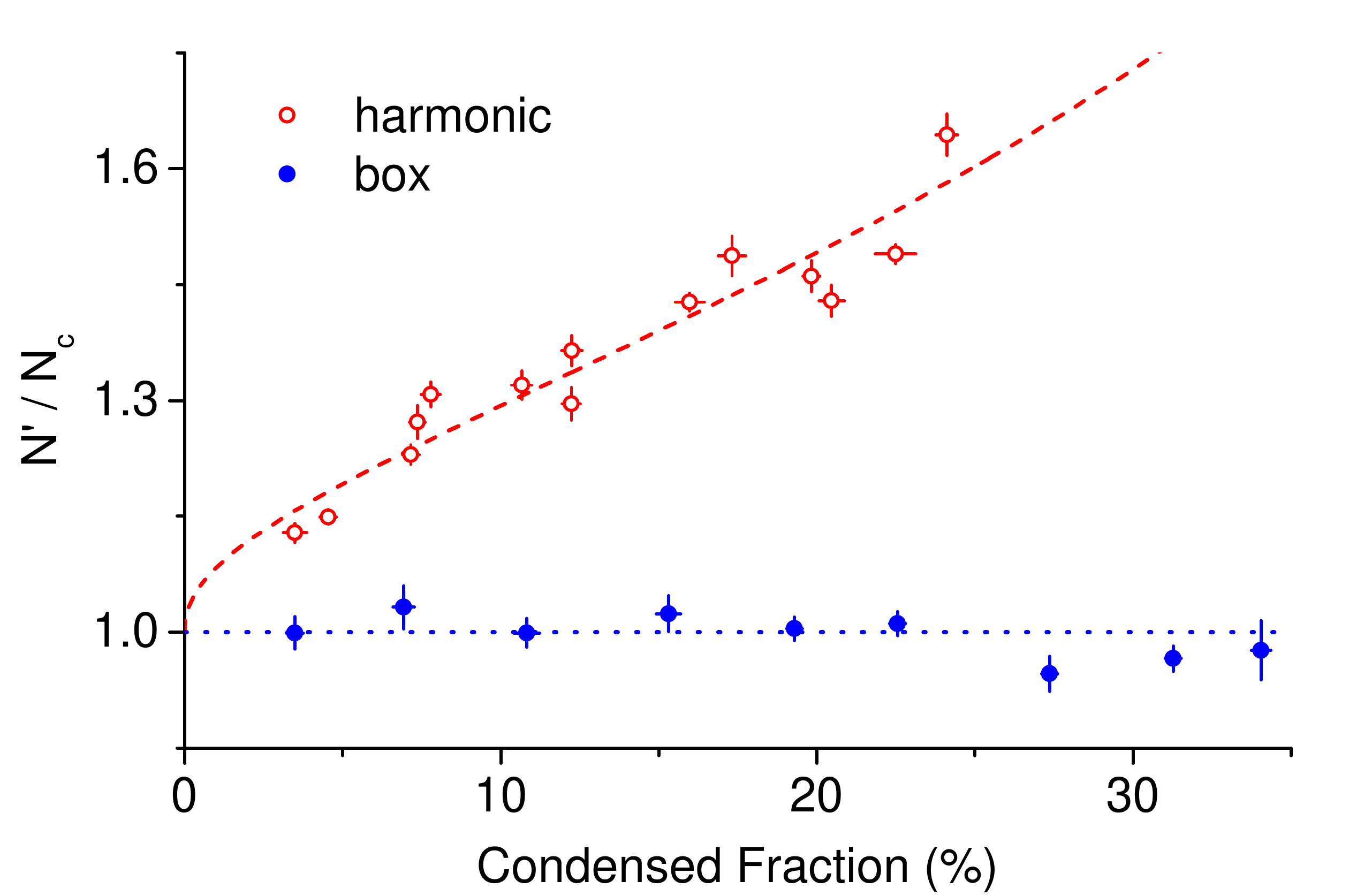}
\caption{(color online) Saturation of the thermal component in a partially condensed gas. In the box trap, the gas follows the ideal-gas prediction, $N' = N_c$, whereas in the harmonic trap the thermal component is strongly non-saturated. Dashed red line shows the theoretical prediction for the harmonic trap. (For the harmonic-trap data $T \approx 110$~nK and $N_c \approx 65\times 10^3$.)}
\label{fig:saturation}
\end{figure}

We now turn to the study of partially condensed clouds.
In Fig.~\ref{fig:saturation} we show the evolution of the thermal atom number as $N$ is increased beyond $N_c$ and a BEC forms. In Einstein's ideal-gas picture of condensation, $N'$ saturates at the critical point and can never exceed $N_c$. 
In a harmonically trapped gas this picture is strongly violated even for weak interactions, such as in $^{87}$Rb, and can be recovered only by extrapolating to the strictly non-interacting limit~\cite{Tammuz:2011}, where direct measurements cannot be performed due to absence of thermal equilibrium~\cite{Gaunt:2013b}. This strong deviation from ideal-gas behaviour  arises due to an interplay of interactions and the non-uniformity of the gas, and can be explained using a mean-field theory that does not require violation of the saturation picture at the level of the local thermal density~\cite{Tammuz:2011,Smith:2013}.

\begin{figure*} [tbp]
\includegraphics[width=\textwidth]{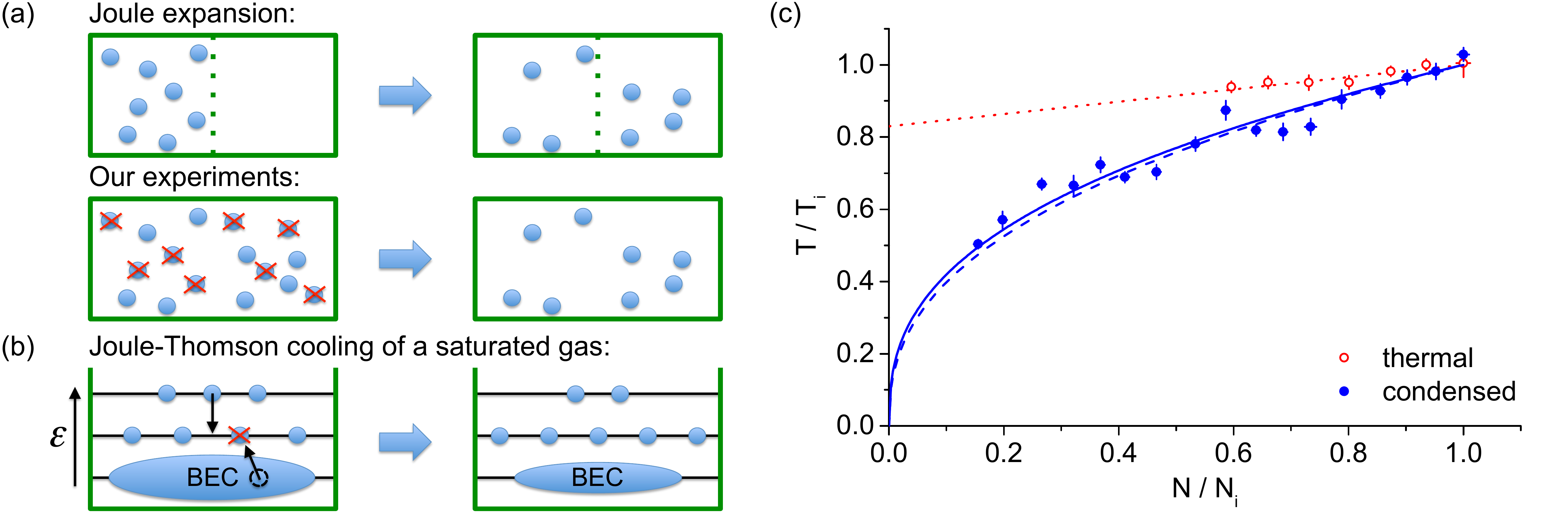}
\caption{(color online) Quantum Joule-Thomson effect. 
(a) Isoenthalpic rarefaction. 
In a conventional JT process $V$ is increased, whereas in our experiments $N$ is reduced, but for an ideal gas 
both processes conserve specific enthalpy and are thermodynamically equivalent.
(b) Microscopic origin of JT cooling in a saturated quantum gas. Removal of a thermal atom requires a zero-energy particle to come out of the BEC in order to maintain $N'=N_c$, while energy conservation requires redistribution of atoms between energy levels. Cooling is seen in the change of the relative populations of the excited states. In this cartoon we neglect the fact that $N_c$ also slowly decreases.
(c) Measurements with a partially condensed and a thermal cloud. Solid and dashed blue line are predictions of Eq.~(\ref{eq:JT}) with $\alpha = 1.65$ and $\alpha = 3/2$, respectively. Dotted red line is a numerical calculation (see text).}
\label{fig:JT}
\end{figure*}

In our experiments, $N'$ is essentially equivalent to the thermal density. In Fig.~\ref{fig:saturation} we see that in this case $N'$ indeed remains constant as the condensed fraction, $N_0/N$, is increased at constant temperature ($45 \pm 1$~nK)~\cite{Negative_saturation}.
For comparison, we also show analogous measurements for the harmonic trap from which our box trap is loaded. Here the non-saturation of $N'$ is prominent and agrees with the prediction based on Refs.~\cite{Tammuz:2011,Smith:2013}, without any free parameters.

Finally, we study the free long-time evolution of a gas held in the box trap, and observe the quantum Joule-Thomson effect  (Fig.~\ref{fig:JT}). 
We prepare clouds at $T_i \approx 45$~nK in a trap of depth $U_0$, where evaporation is negligible.
As the gas is held in the trap, atoms are slowly removed from the cloud, with an exponential timescale $\tau \approx 10$~s, by collisions with the background gas in the vacuum chamber. Due to low atomic density ($< 5\times 10^{12}$~cm$^{-3}$), the three-body recombination rate is negligible~\cite{Soding:1999}. As $N$ decays, the elastic collision rate (among the trapped atoms) remains sufficiently high for the gas to continuously re-equilibrate on a timescale  $\tau_{\rm eq} \lesssim 2$~s~$\ll \tau$~\cite{Monroe:1993}. 

Crucially, collisions with the background gas are independent of the energy of the trapped atoms, so the average energy per trapped particle, $E/N$, remains constant. 
Within the ideal-gas approximation, pressure is simply given by the energy density, so enthalpy, $H$, is simply proportional to $E$, and the specific enthalpy, $h = H/N$, is a conserved quantity.
The decay of the cloud is thus equivalent to the Joule-Thomson isoenthalpic rarefaction, as depicted in Fig.~\ref{fig:JT}(a). For simplicity, here we illustrate free Joule expansion into vacuum; in the JT process rarefaction is driven by throttling, but for an ideal gas both processes are isoenthalpic.

In a {\it classical} ideal gas $h$ depends only on $T$, so temperature cannot change in an isoenthalpic process. 
This is however not true for a quantum degenerate ideal gas~\cite{Kothari:1937,Pethick:2002}. In a partially condensed cloud only thermal atoms contribute to the total energy, so for a saturated gas $E \propto N' T = N_c T$. 
More precisely, $E = \alpha [\zeta(\alpha+1)/\zeta(\alpha)] N_c \kB T$~\cite{Pethick:2002}, where $\zeta$ is the Riemann zeta function.
From Eq.~(\ref{eq:Nc_trap}), keeping $E/N$ constant implies cooling according to
\begin{equation}
 T \propto N^{1/(\alpha + 1)} \, .
\label{eq:JT}
\end{equation}
The microscopic origin of this cooling is qualitatively illustrated in Fig.~\ref{fig:JT}(b). Removal of any thermal atom requires a zero-energy particle to come out of the BEC in order to maintain the saturation of the thermal component. Energy conservation then requires simultaneous redistribution of atoms between energy levels. The net cooling is seen in the change of the relative populations of the excited states. Note that direct removal of BEC atoms does not change the temperature, and also that in this simplified cartoon we neglect the fact that $N_c(T)$ gradually decreases as $T$ drops.

In Fig.~\ref{fig:JT}(c) we show the evolution of $T$ with decaying $N$, with both quantities scaled to their initial values. We show data for a partially condensed gas, with $N_i \approx 1.7 \, N_c (T_i)$, and a thermal cloud, with $N_i \approx 0.6 \, N_c(T_i)$. Note that in both cases we measure down to approximately the same final $N$ value, $\approx 0.3\, N_c(T_i)$. At this point $N_0$ in the initially condensed sample vanishes, and also below this $N$ our temperature fits become less reliable.

In the (partially) condensed sample we observe a drop in $T$ by a factor of two, i.e. by $\approx 22$~nK. Meanwhile, the  interaction energy per particle is always smaller than $(8 \pi \hbar^2 a/m) N_i/V$~\cite{Pethick:2002}, corresponding to $<4$~nK. The change in $T$ therefore must predominantly be a purely quantum-statistical, rather than an interaction effect~\cite{interactions}.

Indeed, Eq.~(\ref{eq:JT}) fits the data very well. We show predictions for $\alpha = 1.65$ (solid blue line) and $\alpha=3/2$ (dashed blue line). The two are almost indistinguishable, reaffirming that the behaviour of our clouds is very close to that of a perfectly homogeneous gas.

Within the constant-$V$ approximation, our measurements are directly related to the Joule-Thomson coefficient, $\mJT = (\partial T/\partial P)_h$, according to $\mJT = A (\partial T/\partial N)_h$, with constant $A = (3/2) VN/E$. More conveniently expressed, the dashed blue line in Fig.~\ref{fig:JT}(c) corresponds to~\cite{Pethick:2002}
\begin{equation}
\mJT = \frac{2}{5\, \zeta(5/2)} \frac{\lambda^3}{\kB} \, .
\label{eq:QJT}
\end{equation}
At our lowest temperatures $\mJT \approx 4 \times 10^9$~K/bar, about ten orders of magnitude larger than observed in classical gases. 

To understand the origins of this enhancement, on dimensional grounds we write $\mJT = \cJT T/P$, so $\cJT$ is dimensionless and $T/P \sim 1/(n' \, \kB)$, where $n'$ is the thermal particle density. The value of $\cJT$ depends on the equation of state. In the ideal gas, $\cJT = 0$ in the classical limit, while $\cJT = 1/(\alpha+1)$ for a saturated Bose gas~\cite{Pethick:2002}. Interaction corrections to $\cJT$ are essentially given by the ratio of interaction and thermal energy.
%which can for a classical gas be estimated using van der Waals parameters.  
In our case these small corrections are unimportant and $\cJT \approx 2/5$, but for a classical gas they are the only origin of the JT effect and typically $\cJT\lesssim 10^{-3}$~\cite{vdW}. The other large enhancement of $\mJT$ comes from the fact that in a saturated gas $n' \sim \lambda^{-3} \sim T^{3/2}$ decreases with temperature, and our lowest $n'$ is $\sim 10^{7}$ times smaller than at ambient temperature and pressure.
Combining the differences in $\cJT$ and $1/n'$, we recover the $\sim 10^{10}$-fold increase in $\mJT$ compared to classical gases. It is also interesting to note that $\mJT$ in Eq.~(\ref{eq:QJT}) explicitly vanishes in the classical limit, $\hbar \rightarrow 0$, reiterating that this is a purely quantum effect.

Compared to the partially condensed gas, in the thermal sample $T$ remains almost constant, but we do discern a slight cooling effect. This is indeed expected for a degenerate non-condensed gas~\cite{Kothari:1937}, with a chemical potential $\mu \gtrsim - \kB T$. 
Degeneracy preferentially enhances occupation of low energy states, with $\varepsilon < |\mu|$ (e.g., momentum distribution is more ``peaky" than a Gaussian~\cite{Gaunt:2013,Pethick:2002}), so $E/N$ is lower than the classical value $\alpha \kB T$. Between the classical limit and the critical point, $\eta \equiv E/(\alpha N \kB T)$ decreases from 1 to $\zeta(\alpha+1)/\zeta(\alpha)$~\cite{Pethick:2002}.
Hence, if $E/N$ is kept constant while $\eta$ grows as the gas becomes less degenerate due to $N$-decay, $T$ must slightly decrease. The dotted red line in Fig.~\ref{fig:JT}(c) shows a numerical calculation of this effect for our $N_i/N_c(T_i)$, and again fits the data very well.

Similar theoretical analysis applies to an ideal harmonically trapped gas, but in that case $\alpha=3$, implying weaker cooling in both condensed clouds and degenerate thermal samples; see Eq.~(\ref{eq:JT}) and note that $\zeta(4)/\zeta(3) \approx 0.9$, as compared to $\zeta(5/2)/\zeta(3/2) \approx 0.5$. Moreover, in practice harmonically trapped gases strongly deviate from this picture, because $N'$ is not saturated~\cite{Tammuz:2011}, much higher typical BEC densities enhance the role of interactions, and three-body recombination continuously increases $E/N$.

In conclusion, we have characterised the critical point for Bose-Einstein condensation in a quasi-homogeneous weakly interacting gas, and demonstrated two textbook-like quantum-statistical phenomena that highlight qualitative differences between uniform and harmonically trapped degenerate gases. In future work, combining our methods with stronger interactions (employing a Feshbach resonance~\cite{Chin:2010}) and flexible shaping of box-like traps~\cite{Gaunt:2012}, should allow further studies of novel thermomechanical effects~\cite{Papoular:2012}. 

We thank Richard Fletcher and Jean Dalibard for critical reading of the manuscript.
This work was supported by EPSRC (Grant No. EP/K003615/1), AFOSR, ARO and DARPA OLE.
R.P.S. acknowledges support from the Royal Society.
N.N. acknowledges support from Trinity College, Cambridge.

%\cs $\lambda = 760$nm and $n_c^0 = 2.612/\lambda^3 = 5.95 \mu m^{-3}$ at 60 nK and $8\pi \hbar^2 a/m = 0.7 nK\,\mu m^3$ and $\pi 15^2 \, 60 = 42411$\cs

%\bibliography{ThermoUniform}

\begin{thebibliography}{10}
\bibitem{Nc}
Since $N'$ saturates for $N>N_c$, as shown in Fig.~\ref{fig:saturation}, associating the critical point with a small but non-zero $N_0$ does not introduce any errors.
\bibitem{vdW}
For a classical gas, $\cJT$ can be estimated, to within an order of magnitude, using the van der Waals approximation for the  equation of state.
\end{thebibliography}
%\bibliographystyle{prsty}

\end{document}